\documentclass[sigconf, nonacm]{acmart}

\usepackage[]{hyperref}
\usepackage{textcomp}
\usepackage{xcolor}
\usepackage{subcaption}
\usepackage[ruled, linesnumbered]{algorithm2e}
\usepackage{color}
\usepackage{multirow}
\usepackage{pifont}
\newcommand*\circled[1]{\raisebox{.5pt}{\textcircled{\raisebox{-.5pt} {\footnotesize{\textbf{#1}}}}}}

\AtBeginDocument{%
  }

\copyrightyear{2025} 
\acmYear{2025} 
\setcopyright{cc}
\setcctype{by}
\acmConference[MICRO '25]{58th IEEE/ACM International Symposium on Microarchitecture}{October 18--22, 2025}{Seoul, Republic of Korea}
\acmBooktitle{58th IEEE/ACM International Symposium on Microarchitecture (MICRO '25), October 18--22, 2025, Seoul, Republic of Korea}
\acmDOI{10.1145/3725843.3756110}
\acmISBN{979-8-4007-1573-0/2025/10}

\begin{document}


\title{A TRRIP Down Memory Lane: Temperature-Based Re-Reference Interval Prediction For Instruction Caching}
\author{Henry Kao}
\affiliation{
  \institution{Huawei Technologies Canada}
  \city{Toronto}
  \state{ON}
  \country{Canada}
}
\email{henry.kao1@huawei.com}

\author{Nikhil Sreekumar}
\affiliation{
  \institution{Huawei Technologies Canada}
  \city{Toronto}
  \state{ON}
  \country{Canada}
}
\email{nikhil.sreekumar@huawei.com}

\author{Prabhdeep Singh Soni}
\affiliation{
  \institution{Huawei Technologies Canada}
  \city{Toronto}
  \state{ON}
  \country{Canada}
}
\email{prabhdeep.singh.soni3@huawei.com}

\author{Ali Sedaghati}
\affiliation{
  \institution{Huawei Technologies Canada}
  \city{Toronto}
  \state{ON}
  \country{Canada}
}
\email{ali.8296571@huawei.com}

\author{Fang Su}
\affiliation{
  \institution{Huawei Technologies}
  \city{Beijing}
  \country{China}
}
\email{fang.su@huawei.com}

\author{Bryan Chan}
\affiliation{
  \institution{Huawei Technologies Canada}
  \city{Toronto}
  \state{ON}
  \country{Canada}
}
\email{bryan.chan@huawei.com}

\author{Maziar Goudarzi}
\affiliation{
  \institution{Huawei Technologies Canada}
  \city{Toronto}
  \state{ON}
  \country{Canada}
}
\email{maziar.goudarzi@huawei.com}

\author{Reza Azimi}
\affiliation{
  \institution{Huawei Technologies Canada}
  \city{Toronto}
  \state{ON}
  \country{Canada}
}
\email{reza.azimi1@huawei.com}

\renewcommand{\shortauthors}{H. Kao, N. Sreekumar, P. S. Soni, A. Sedaghati, F. Su, B. Chan, M. Goudarzi and R. Azimi}





\begin{abstract}


Modern mobile CPU software pose challenges for conventional instruction cache replacement policies due to their complex runtime behavior causing high reuse distance between executions of the same instruction. Mobile code commonly suffers from large amounts of stalls in the CPU frontend and thus starvation of the rest of the CPU resources. Complexity of these applications and their code footprint are projected to grow at a rate faster than available on-chip memory due to power and area constraints, making conventional hardware-centric methods for managing instruction caches to be inadequate.
We present a novel software-hardware co-design approach called TRRIP (Temperature-based Re-Reference Interval Prediction) that enables the compiler to analyze, classify, and transform code based on "temperature" (hot/cold), and to provide the hardware with a summary of code temperature information through a well-defined OS interface based on using code page attributes. TRRIP's lightweight hardware extension employs code temperature attributes to optimize the instruction cache replacement policy resulting in the eviction rate reduction of hot code.
TRRIP is designed to be practical and adoptable in real mobile systems that have strict feature requirements on both the software and hardware components.
TRRIP can reduce the L2 MPKI for instructions by 26.5\% resulting in geomean speedup of 3.9\%, on top of RRIP cache replacement running mobile code already optimized using PGO.


\end{abstract}


\keywords{Compiler, Profile-Guided Optimization, Microarchitecture, Cache Replacement, Operating System}

\maketitle

\section{Introduction}

Modern applications in mobile workloads are becoming increasingly complex with layered software architecture consisting of many libraries resulting in deep call stacks, high-level language constructs, intricate control flow, and large instruction footprints \cite{ioscodesize, bytedance}.
Cache replacement policies struggle with the demands of modern mobile workloads as they are only able to observe a relatively small window of the program execution to decide which line to keep in or out. The overall behavior of the program (i.e., memory access characteristics) can become difficult to detect and track for pure hardware techniques. Tracking enough runtime information requires additional on-chip memory as in the state-of-the-art techniques improving cache re-reference interval prediction~\cite{hawkeye, ship, mockingjay}, instruction prefetchers that operate as runahead~\cite{fdip,pdip}, and record-and-replay~\cite{pif,mana} -- all aiming to reduce CPU frontend stalls at the expense of power and area.
Thus conventional cache hierarchies perform sub-optimally under these complicated workloads due to both convoluted instruction access patterns and large code footprints. Increasing cache size or adding new predictor tables is unsustainable since power and area constraints prevent on-chip storage from growing at a rate matching the growth of code -- upward of 2.7x a year~\cite{ioscodesize}.

Numerous prior studies have addressed the frontend bottleneck on datacenter machines~\cite{twig, emissary, ispy, ripple, asmdb, profilingWSC}, but few analyses or solutions have been proposed on mobile platforms with stricter restrictions on power consumption. We notice that frontend bottlenecks also occur on different software components of mobile platforms. Profile-guided optimization (PGO) is a widely employed technique~\cite{autofdo,bolt,csspgo,propeller} to obtain further performance benefits, often in the form of code layout optimization to reduce frontend stalls. Profile data provides additional information such as code \textit{temperature} -- which parts of the code take up a significant portion of the total execution of the application (\textit{hot} code), which parts of code are rarely, or never executed (\textit{cold} code), and which parts of code are \textit{warm} (neither \textit{hot} nor \textit{cold}). PGO includes traditional compiler optimizations (e.g., function inlining, function reordering, block placement), as well as novel software prefetching techniques~\cite{ispy, asmdb}. We have observed considerable reduction in cycles in the most frequently used mobile system software when compiled with PGO. PGO is now integrated into the compilation of commercial mobile system software components as it can squeeze more performance out of a processor without any hardware changes. PGO and the profiling information provides a global end-to-end characterization of the application -- additional intelligence which can complement hardware techniques that struggle to track enough runtime data. We observe that the most executed (\textit{hot}) code still shows adverse instruction caching behavior even after PGO; \textit{Hot} instruction cache lines exhibit high re-reference intervals in which conventional cache policies struggle with resulting in CPU frontend stalls.

Our key contribution is a compiler, OS (Operating System) and hardware co-design technique called TRRIP (Temperature-Based Re-Reference Interval Prediction) to form a lightweight end-to-end solution that tackles frontend stalls in modern workloads, and is designed to be practical and adoptable in real mobile systems. TRRIP passes code temperature information to hardware using a lightweight interface by tagging code with temperature bits at page-level granularity to make a more informed prediction of re-reference intervals. TRRIP's cache replacement policy makes use of code temperature to prioritize keeping \textit{hot} instruction lines in the cache in order to prevent cache misses, and hence frontend stalls on the most executed code. TRRIP incurs negligible power and area costs by utilizing PGO to do offline re-reference analysis. Our contributions are:

\begin{itemize}
    \item Characterization of real mobile system software and PGO'd proxy applications showing frontend bottleneck problem and memory characteristics of the most executed (\textit{hot}) code.
    \item TRRIP: a software-hardware co-design cache replacement technique that reduces instruction cache misses and hence CPU frontend stalls through prioritizing \textit{hot} code.
    \item An evaluation of TRRIP demonstrating geomean 3.9\% speedup, via a reduction of L2 instruction MPKI by 26.5\% on applications that have already been optimized using PGO, with minimal changes to the cache replacement policy.
\end{itemize}

\section{Background \& Motivation}

\subsection{Frontend Bottleneck in the Mobile System}
\label{sec:frontend-bottleneck}

The system software on mobile platforms form the core functionality of the device, which include shared libraries that implement user interface behavior, graphics rendering, hardware abstraction interfaces, interprocess communication, language runtimes, as well as interpreters and JIT (Just in Time)/AOT (Ahead of Time) compilers. All applications will call/use system software components during their execution, thus system software performance optimizations do not only pertain to specific applications; their effect will be observed system-wide spanning across all applications. The system software is primarily written in C/C++ to provide high performance. We have observed that the system software components exhibit frontend bottlenecks due to instruction cache misses. Instrumentation PGO along with the associated code layout optimizations have shown considerable performance improvement, and is now integrated into the compilation flow of mobile system software, however the frontend bottleneck still appears to be the most prevalent bottleneck even with PGO applied.

\begin{figure}[!t]
\centering
\includegraphics[width = 1.0\columnwidth]{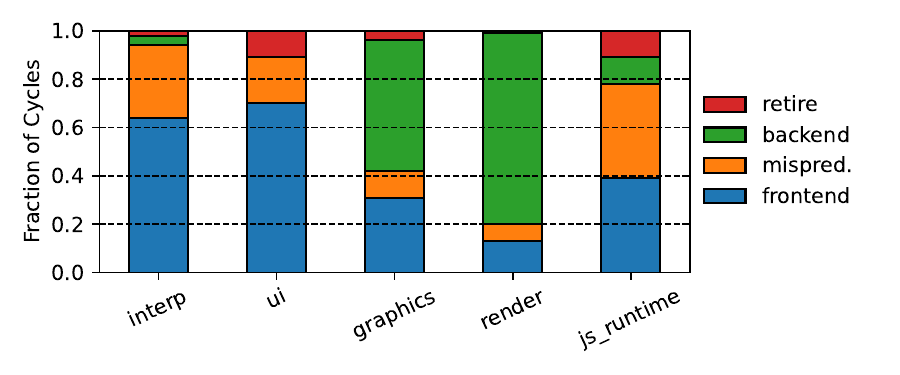}
\caption{Top-Down breakdown of hottest system software components which includes code interpreter (\texttt{interp}), and shared libraries for a user-interface framework (\texttt{ui}), graphics (\texttt{graphics}), rendering (\texttt{render}), and JavaScript runtime (\texttt{js\_runtime}).}
\label{fig:top-down-shared-libs}
\end{figure}

Figure~\ref{fig:top-down-shared-libs} shows a Top-Down~\cite{top-down-method} breakdown of cycles for the hottest 
components in OpenHarmony 5~\cite{openharmony} that are compiled with PGO. Top-Down methodology samples PMU counters to infer where the bottleneck occurs in the processor; in the \texttt{frontend} due to instruction cache misses, mispredictions (\texttt{mispred.}), or stalls in the CPU \texttt{backend} (includes core and memory side stalls). Cycles can also be usefully spent retiring instructions (\texttt{retire}). The mobile system software was profiled using a photo viewing application running on a Huawei Mate 60 Pro~\cite{mate60pro} with a Kirin 9000S chipset.
The hottest system software that are profiled still show a considerable amount of frontend bottlenecks even with PGO applied.

\subsection{Motivating Through Proxy Benchmarks}
\label{sec:proxy-benchmarks}

The system software for mobile platforms are tightly coupled to the rest of the complicated software stack which includes proprietary and third-party code which are difficult to extract into a standalone benchmark for evaluation. We gather several applications from publicly available repositories that act as a proxy for components in mobile system software. The chosen benchmarks are C/C++ based for high baseline performance, and try to mimic mobile system functionalities including code generators (e.g., interpreters, JIT and AOT optimizers) as well as utilities in the shared system libraries.
Section~\ref{sec:benchmarks} further details the proxy benchmarks and compilation procedure.

\subsection{PGO Cannot Fix The Frontend Bound Issue}
\label{sec:pgo-motivation}
\begin{figure*}[!th]
\centering
\includegraphics[width = 2\columnwidth]{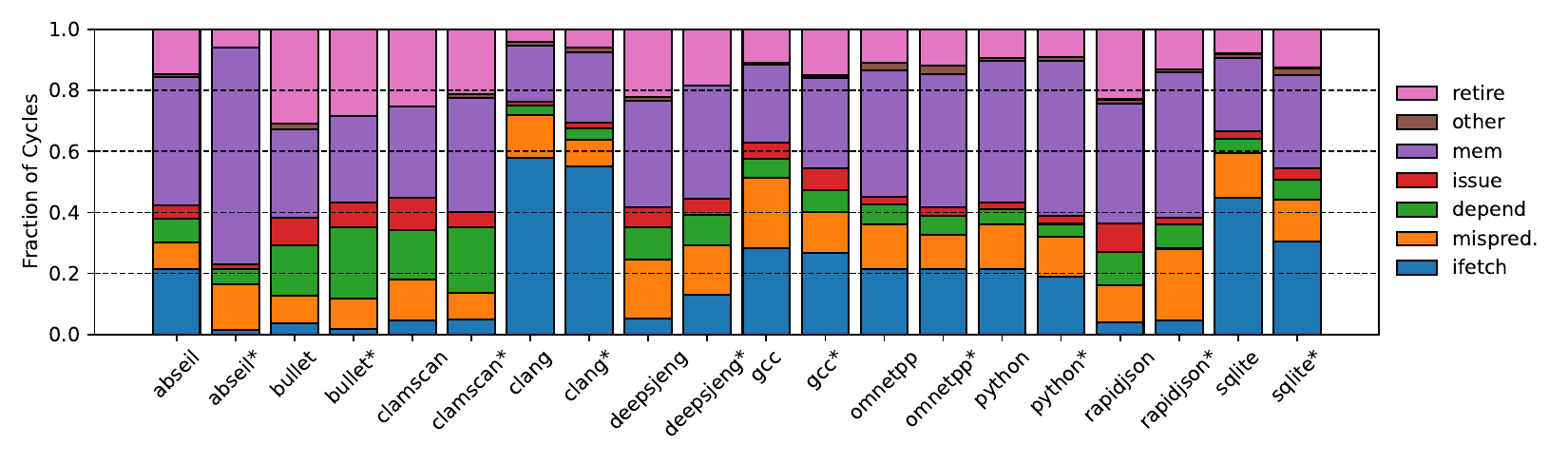}
\caption{Top-Down profiles of proxy mobile benchmarks. Non-PGO compile and PGO compile (marked with ``*'') are shown. Cycles spent doing useful computation is shown as \texttt{retire}. The rest are cycles lost due to frontend stalls from instruction cache misses (\texttt{ifetch}), branch misprediction (\texttt{mispred.}), data dependencies (\texttt{depend}), saturated issue queues (\texttt{issue}), and backend stalls waiting for data from caches and main memory (\texttt{mem}).}
\label{fig:cpi-stack}
\end{figure*}

Figure~\ref{fig:cpi-stack} is a Top-Down analysis of the proxy benchmarks on a simulation platform detailed in Section~\ref{sec:simulator}. Fraction of cycles spent doing useful work/computation is shown as (\texttt{retire}) or stalled in the different stages of the CPU. Stalled cycles are collected from instruction fetch (\texttt{ifetch}), branch mispredictions (\texttt{mispred.}), data dependencies (\texttt{depend}), full issue queues (\texttt{issue}), and CPU backend data access due to latencies in accessing caches and DRAM (\texttt{mem}). Stalls not accounted for in the aforementioned stages are classified as \texttt{other}. The Top-Down profiles are collected for each proxy benchmark twice, once without PGO, and once with PGO. The PGO'd version is marked with an ``*''. Many of the non-PGO'd benchmarks (\textit{abseil}, \textit{clang}, \textit{gcc}, \textit{omnetpp}, \textit{python}, \textit{sqlite}) show over 20\% of the stall cycles coming from \texttt{ifetch} -- due to misses in the instruction caches. The effects of PGO can be seen by comparing the non-PGO versus the PGO'd Top-Down profile for the same benchmark (e.g., \textit{sqlite} versus \textit{sqlite*}). In general, the increase in fraction of \texttt{retire} means that the benchmark spends more cycles doing useful work instead of stalling. Much of the gain in \texttt{retire} comes from PGO-enabled compiler optimizations reducing stalls in the CPU frontend including \texttt{ifetch} stalls and \texttt{branch} stalls. The various code layout optimizations under PGO improve code spatial locality which reduces cache misses on instruction memory, and also creates better BB (Basic Block) layouts (by creating more fall-through blocks) to mitigate branch mispredictions. However the benchmarks still show considerable amount of cycles stalled on \texttt{ifetch} even after using PGO. Stalls in the frontend are problematic as the subsequent stages of the CPU pipeline are starved of work. 
In practice, PGO does not always guarantee performance gains. Occasional degradations may happen (as in the case of deepsjeng and rapidjson) mainly due to the differences between the profile data used for the PGO and the actual behavior of the application\footnote{Using different input sets between PGO profile collection and evaluation is a common practice in the industry to maintain robust and representative profiles.}.

\subsection{Hot Code Shows High Reuse Distance}

\begin{figure}[!t]
\centering
\includegraphics[width = 1\columnwidth]{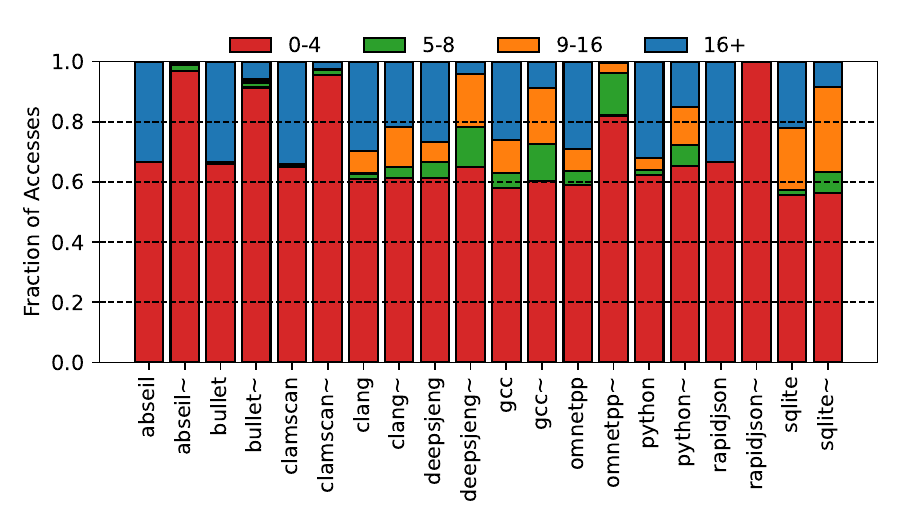}
\caption{Reuse distance distribution of hot cache lines measured in the L2 cache. Reuse is measured as the number of unique cache lines seen between two subsequent access of the same line for one given cache set. Applications post-fixed with ``$\sim$'' measures reuse distance only counting \textit{hot} unique cache lines seen between two subsequent access of the same line for one given cache set.}
\label{fig:l2-reuse-cdf}
\end{figure}

Why do applications still show considerable frontend stalls even though PGO is applied? PGO itself is able to classify code \textit{temperature} (i.e., \textit{hot}, \textit{warm} and \textit{cold}) by calculating the contribution of a region of code to the total amount of execution time. Code is classified as \textit{hot} if it contributes a large portion to the total. Code is \textit{cold} if it only contributes a negligible portion to the total. \textit{Warm} code can be considered as neither \textit{hot} nor \textit{cold}. We look at the \textit{hot} code to find clues to answer the question why we still get instruction fetch stalls even after PGO, since the \textit{hot} code generally makes up most of an benchmark's execution. PGO in the compiler itself already makes a best effort to improve spatial locality of \textit{hot} code, so we inspect the temporal locality instead by measuring reuse distance of the instruction cache lines corresponding to the \textit{hot}. We measure reuse on a cache set granularity as the number of unique cache lines (both instruction and data) seen between two subsequent access of the same line. Modern high performance CPUs generally tolerate L1-I cache misses well due to the lower miss penalty~\cite{emissary}. Misses in L2 incur considerably higher costs due to the increasing access latencies in downstream caches and main memory which are often off-chip, thus we target the L2 for our reuse distance measurements. Figure~\ref{fig:l2-reuse-cdf} shows a breakdown of reuse distances of \textit{hot} instruction cache lines at the unified L2 cache level.

The reuse distance is measured and shown for each benchmark in two ways; (1) a base measurement where all unique lines are counted between subsequent accesses of a \textit{hot} line and (2) and an optimistic measurement where only unique \textit{hot} lines are counted between subsequent accesses of a \textit{hot} line (post-fixed with ``$\sim$''). The former measures true temporal locality of \textit{hot} lines in the benchmark and the latter measures temporal locality of \textit{hot} lines in the absence of non-\textit{hot} lines (i.e., \textit{warm}, \textit{cold} and data lines).  Most of the \textit{hot} instruction lines show high temporal locality in the base measurements since short reuse distances, from 0 to 4, makes up $\sim$60\% of the accesses. Assuming a cache with a conventional replacement policy that inserts new lines as MRU (Most Recently Used) (i.e., LRU replacement), or promotes existing lines to MRU (e.g., RRIP replacement~\cite{rrip}), a cache with 4-way set associativity should be able to keep most of this portion of \textit{hot} lines in the cache without misses. Up to 8-way set associativity should be able to keep most of the \textit{hot} lines that make up 0-4 and 5-8 reuse distance in the distribution without incurring substantial misses. Hot cache lines that have reuse distance of 9 or greater will be evicted out of an 8-way set associative in conventional replacement policies. This measurement explains why these mobile benchmarks optimized using PGO still exhibit high frontend stalls. A large portion of the code that takes up a large portion of the benchmark's instruction count, \textit{hot} code, are evicted out of the cache before its next use due to poor temporal locality (i.e., high reuse distances). A considerable number of \textit{hot} code evictions are caused by allocating non-\textit{hot} lines to the cache set seen by comparing the base and optimistic reuse distance measurements. A 16-way cache would be able to store almost all \textit{hot} lines in \textit{abseil$\sim$}, \textit{clamscan$\sim$}, \textit{omnetpp$\sim$}, and \textit{rapidjson$\sim$} if not for evictions caused by non-\textit{hot} lines. A method to prioritize \textit{hot} lines and de-prioritize non-\textit{hot} lines in cache sets would mitigate evictions on code lines most executed by the program.

Conventional and state-of-the-art cache replacement policies aim to keep frequently used memory in the caches for longer, and to keep infrequently used memory out of the caches. RRIP (Re-Reference Interval Prediction)~\cite{rrip} pessimistically assumes that new cache lines will have long reuse distances -- a distant re-reference prediction, and insert them into the cache at a lower priority. Only hits to cache lines upgrade the re-reference prediction to a short re-reference (i.e., short reuse distance) and upgrades the line to a higher priority. RRIP-based policies generally perform better than LRU~\cite{rrip, clip, ship}, albeit still being limited to seeing only short phases of the memory characteristics of the code due to hardware budget and the high costs of tracking detailed and deep history runtime behavior. Application behaviors spanning over long duration can only be realistically captured using software profiling techniques as in the case of PGO. Software profiling is able to classify code \textit{temperature} for substantially longer execution windows -- not only short phases as in hardware-centric techniques. We want to mitigate misses to \textit{hot} code since it makes up a large percentile of the program's total execution. Following Amdahl's law~\cite{amdahl}, any beneficial impact to \textit{hot} code here should have the highest impact overall. Cache replacement policies would benefit from knowing the \textit{temperature} of instruction memory when deciding how to prioritize keeping it in the cache. Ideally the hotter the code, the longer the corresponding cache lines should stay in the cache. However this information can only be sustainably obtained using software mechanisms. We propose a software-hardware co-design method, leveraging PGO, to pass code \textit{temperature} to the caches in order to reduce frontend stalls for these complicated modern applications.

\begin{figure*}[!t]
\centering
\includegraphics[width = 1.85\columnwidth]{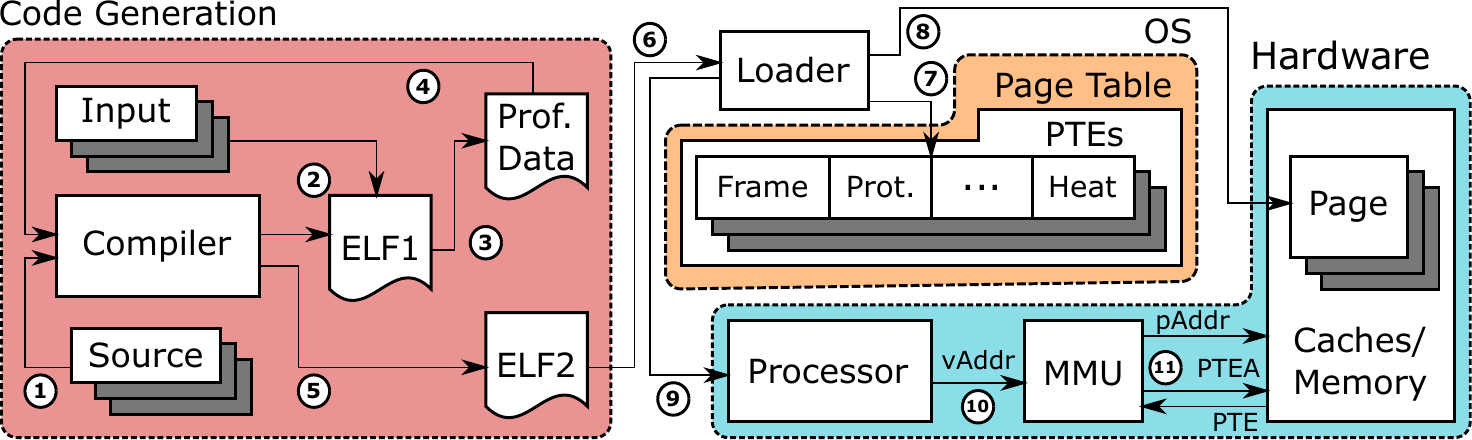}
\caption{Co-designed components and interfaces for TRRIP cache replacement.}
\label{fig:TRRIP-diagram}
\end{figure*}

\section{TRRIP: Temperature-Based Re-Reference Interval Prediction}
\label{sec:design}
TRRIP is a software-hardware co-designed cache replacement policy that utilizes PGO to determine the insertion priority of instruction cache lines in the cache. We have observed several pain-points preventing co-designed ideas from being implemented in real platforms. We consider several key points in our design philosophy to maximize chances of TRRIP's adoption in mobile systems: 

\textbf{No additions/modifications to ISA:} Commercial architecture licenses have restrictions on modifying the ISA which imposes a barrier to adoption on prior co-design methods proposing new instructions to pass software hints to the hardware~\cite{thermometer, ripple, ispy}. Mobile devices also contain heterogeneous CPUs which implement different microarchitectures -- all of which need to obey the same ISA. Thus it is insufficient to only implement an ISA modification to only one microarchitecture in the system. All microarchitectures have to be updated to support the ISA change, creating a logistically complicated design/implementation plan, as each microarchitecture might be handled by different company teams. We make no additions/modifications to the ISA in order to promote easy adoption.

\textbf{Practical Profiling and Compilation:} Previous co-design methods propose collection of execution traces for more accurate offline analysis. Mobile device system monitors abort tasks when performance anomalies occur, for example, aborting frame rendering if it is running past its scheduled deadline (known as frame drop or jank). Collecting profiles with enough coverage of total application behavior can have high runtime and storage overheads, causing the system to abort tasks prematurely interrupting/distorting trace collection. Prior co-design techniques also proposes detailed offline analysis~\cite{thermometer, ripple, ispy}. Introducing high compilation time, compilation complexity, and increased code size are often real hindrances to adoption of software-only optimizations, let alone co-design optimizations. An example of this is the preference to use lower compile time ThinLTO~\cite{thinlto} versus higher performance full LTO within the industry~\cite{bytedance-codesize, meta-reorder, meta-merge}. We opt to keep the compilation flow as-is, which already includes the collection and maintenance of instrumentation profiles and PGO.

\textbf{Minimal Changes to Hardware:} Hardware power consumption translating to battery life and thermal management is an important optimization metric on mobile platforms. Microarchitecture features that offer the best performance-per-watt are preferred. TRRIP is designed to keep hardware changes to a minimum for easier adoption and to prevent additional power consumption overheads.

The novelty of TRRIP is the amalgam and interface of cross-stack components spanning between compiler, OS, and hardware to create an end-to-end low cost solution addressing CPU front-end bottlenecks.  TRRIP's cache replacement policy decides how to prioritize incoming instruction cache lines based on the temperature value that is sent with memory requests requiring negligible changes to the cache replacement policy, no changes in the ISA, no instruction overhead, no need for additional storage on-chip, and no additional impact to compile time/complexity.

\subsection{Co-Design Overview}

Figure~\ref{fig:TRRIP-diagram} illustrates the various interfaces between co-designed components of TRRIP labeled with numbers to follow the description of the technique. \circled{1} The flow starts with compiling program source code using a PGO-enabled compiler. Note that the compiler may also be binary optimizers (e.g., BOLT~\cite{bolt}, Propeller~\cite{propeller}) where the source could be pre-compiled binaries. \circled{2} The compiler will create the first executable, \texttt{ELF1}. The rest of the paper will assume ELF\footnote{The ELF executable format is predominantly used on Linux and Unix based systems, including mobile or embedded platforms, such as Android.} (Executable and Linkable Format) object files, however the techniques can be applied to other object file formats as well. The executable is run with input sets representative of how it will be used in real-world scenarios, after which ~\circled{3} a profile of the application will be generated. \circled{4} The profile along with the original source of the program \circled{1} is fed back into the PGO-enabled compiler to re-optimized the code. \circled{5} A final executable \texttt{ELF2} is generated. This re-optimized executable uses profiling results \circled{4} to place code into different code sections of the ELF based on the classified temperature.

Executing a program requires it to be loaded into memory first, which is the job of the loader. \circled{6} The loader reads headers and sections in \texttt{ELF2} along with linking and runtime information to map the program to pages in memory. The loader calls functionality of the OS to allocate pages \circled{8} for the executable, and also generate corresponding PTEs (Page Table Entry) \circled{7}, populating each entry with the runtime information obtained from \texttt{ELF2}. The typical information may include access permission flags (e.g., read-only, read-write). 
We modify the loader to read and populate existing additional bits~\cite{cortexa, cortexx, AMD64} allocated in the PTE to store temperature-based information of the code sections in \texttt{ELF2}. 
We choose to set code temperature on page-granularity since we are able to reuse existing commercial processor features (i.e., ARM PBHA~\cite{cortexa}, AVL bits in x86 PTE~\cite{AMD64}) offering implementation-defined PTE bits to be transferred with memory requests. It allows for TRRIP to have a transparent light-weight interface to transfer information between software and hardware with minimal to no additional implementation cost, in contrast to prior art which requires new instructions~\cite{twig, ripple} or modification to the existing ISA~\cite{thermometer}. Storing code temperature in the page tables also mitigates additional logic/storage need in the CPU microarchitecture or caches, as in prior work which increases storage requirements by adding additional bits per cache line~\cite{emissary, ship} to track runtime metadata.

\circled{9} The program executes conventionally after being loaded into memory. 
\circled{10} Instructions of the loaded program are fetched from caches or memory. This may require the translation of the instruction addresses from virtual (\texttt{vAddr}) to physical (\texttt{pAddr}) by the MMU~\circled{11}.
PTE temperature bits in the MMU are read and transferred along with the memory requests to the caches. The replacement policies are augmented to react to the temperature bits in the memory requests to prioritize keeping hot instruction memory in the cache for longer, while keeping cold code out of the cache to reduce CPU frontend stalls due to instruction cache misses.

\subsection{Code Generation \& Optimization}
\label{sec:compiler}
\begin{figure}[!t]
\centering
\includegraphics[width = 0.85\columnwidth]{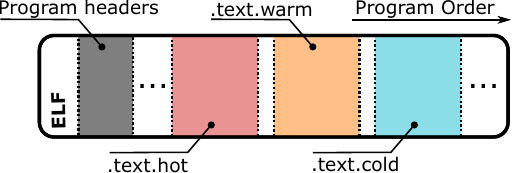}
\caption{ELF layout after PGO showing only the components TRRIP modifies, program headers and \texttt{.text} sections.}
\label{fig:elf-layout}
\end{figure}
TRRIP relies on PGO to classify code based on temperature which guides cache insertion/update policies for more informed re-reference prediction. 
We use and describe our approach using LLVM~\cite{llvm}, an open-source compiler with instrumentation-based PGO as it is the most accurate form of profiling and empirically offers the highest performance improvement~\cite{csspgo} . Note that TRRIP is easily adaptable to any compiler as well as sampling-based PGO~\cite{autofdo}.
Applications are first compiled with instrumentation PGO and executed with representative input sets. Instrumentation counts the execution frequency of BBs to generate a profile. The profile is fed back into the compiler along with the application source code to re-compile for a further optimized executable. 
Mobile system components mostly comprise of shared libraries used by many applications. Shared libraries are also optimized using PGO similar to normal applications, but in the case of shared libraries, the profile data is accumulated through the execution of multiple applications that call code from the same shared library.
There are various compiler optimizations that utilize PGO information, such as function inlining and BB placement, most of which optimize code layout to decrease CPU frontend stalls. We are interested in using profiling information to create discriminate code regions based on code temperature, namely \textit{hot}, \textit{warm}, or \textit{cold}. 
The code temperature calculation is described in detail in Section~\ref{sec:eval-hot-threshold}. 
The temperature-labeled code regions are placed into different code sections when the compiler generates the final ELF as seen in Figure~\ref{fig:elf-layout}. \textit{Hot} code is placed in a single continuous \texttt{.text.hot} section. \textit{Warm} and \textit{cold} code are placed in their own respective sections, separate from other code sections of a different temperature. As a pure code optimization technique, placing \textit{cold} code away from \textit{hot} code already improves instruction caching performance by improving spatial locality of \textit{hot} code.
The ELF program headers contains runtime information needed by the OS before running the application. The runtime information can be specific the different sections of the ELF. ELF headers are modified to include temperature information of the \textit{hot}, \textit{cold} and \textit{warm} code sections.

\subsection{Operating System}
\label{sec:os}

The program loader is responsible in loading an ELF and its dependencies into memory prior to execution. The loader along with the OS generates the page tables and the corresponding PTEs. The loader reads the program header of the ELF to determine the flags associated with the page about to be allocated. These flags are passed to the system call \texttt{mmap} which allocates the pages with the corresponding flags in the program header. \texttt{mmap} also allocates the PTEs for a given section in the ELF and sets its corresponding PTE fields~\cite{linux-pte-helpers}.
Implementation-defined PTE bits (ARM's PBHA feature) already exist in commercial mobile processors~\cite{cortexa, cortexx}. We use these extra bits in the PTEs to track temperature of pages containing code -- requiring no additional implementation/storage cost.

\SetAlFnt{\small}
\begin{algorithm}[ht]
\caption{Re-reference interval prediction policy for TRRIP. Temperature sensitive additions to RRIP are shown in red.}\label{alg:trrip-policy}
\SetKwInOut{Tag}{Tag}
\SetKwInOut{Temp}{Temp}
\Tag{Cache block address tag}
\Temp{Cache block temperature \{hot, warm, cold, none\}}
\vspace{1mm}
\eIf{Tag Hit}{ 
    \tcc{Cache hit, update re-ref. prediction of line}
    \Switch{Temp} {
        \textcolor{red}{\Case{hot}{
            \tcc{TRRIP variant 1 \& 2}
            $\text{Line}_{\text{RRPV}} = \textsc{immediate re-ref.}$\
        }
        \Case{warm $||$ cold} {
            \tcc{TRRIP variant 2 only }
            $\text{Line}_{\text{RRPV}} = \textsc{Max}(\text{Line}_{\text{RRPV}}- 1,\textsc{immediate})$\
        }}
        \Other{
            \tcc{Default behavior}
            $\text{Line}_{\text{RRPV}} = \textsc{immediate re-ref.}$\
        }
    }
}
{
    \tcc{Cache miss, do replacement. Increment re-ref. of all lines until eviction is found} \
    $\text{Line} \gets \textsc{GetEvictionLine()}$\

    \tcc{Insert line and set RRPV value}
    \Switch{Temp} {
        \textcolor{red}{\Case{hot}{
            \tcc{TRRIP variant 1 \& 2}
            $\text{Line}_{\text{RRPV}} = \textsc{immediate re-ref.}$\
        }
        \Case{warm} {
            \tcc{TRRIP variant 2 only}
            $\text{Line}_{\text{RRPV}} = \textsc{near re-ref.}$\
        }}
        \Other{
        \tcc{Default behavior}
            $\text{Line}_{\text{RRPV}} = \textsc{intermediate re-ref.}$\
        }
    }
}
\end{algorithm}

\subsection{Cache Replacement Policy}
\label{sec:hardware}


Algorithm~\ref{alg:trrip-policy} shows the implementation of the TRRIP cache replacement policy, specifically the insertion and update sub-policies. TRRIP's insertion policy utilizes temperature hints from PGO to predict the re-reference behavior of the instruction cache lines and is built upon existing RRIP cache insertion/replacement policies~\cite{rrip}. The temperature information is transferred with memory requests to the cache, along with the memory address, thus no extra storage is needed in the caches. 
The conventional RRIP policy aims to predict the re-reference interval of a cache line within a cache set. RRIP encodes re-reference predictions for each cache line using an RRPV (Re-Reference Prediction Value) which is a fixed number of bits per line.
Lower RRPV values represent more immediate re-reference predictions and thus higher priority it is to be kept in the cache. Highest to lowest re-reference priorities using a 2-bit RRPV example follows as: \texttt{Immediate (RRPV=0)} $>$ \texttt{Near (RRPV=1)} $>$ \texttt{Intermediate (RRPV=2)} $>$ \texttt{Distant (RRPV=3)}. New lines are pessimistically inserted at \texttt{Intermediate} re-reference in baseline RRIP policies (line 22-24). Hits on a cache line immediately set to \texttt{Immediate} re-reference (line 9-11), predicting the next access to the set will likely access the same line. The higher the RRPV value, the more likely the line is a candidate for eviction as it represents increasingly distant re-reference prediction. TRRIP does not modify the eviction mechanism (\texttt{GetEvictionLine}) of RRIP (line 14),  which increments the RRPV of all ways in a set until an RRPV with maximum value is found and selected for replacement. Parts of the algorithm in red are the TRRIP additions to RRIP. The key idea behind TRRIP is to keep the \textit{hot} instruction cache lines in cache for as long as possible to minimize frontend stalls for the most-frequently executed code. New \textit{hot} instruction lines are inserted as \texttt{Immediate} (line 16-18) to prevent premature eviction. Cache hits to \textit{hot} lines follows default behavior and promoted to \texttt{Immediate} (line 3-5). To keep \textit{hot} lines in the highest priority position (\texttt{Immediate}) for longer, \textit{warm} and \textit{cold} instruction lines are not directly promoted to \textit{Immediate} upon a hit, instead the RRPV is conservatively decremented (lines 6-8). Insertion of \textit{warm} lines are inserted as \texttt{Near} re-reference, having higher priority than other lines (i.e, data lines) since we are targeting frontend bound applications, but also lower priority than \textit{hot} lines (lines 19-21). \textit{Cold} instruction lines, by definition, are rarely executed (or never at all), thus having a separate policy just for \textit{cold} lines will not have much end-to-end effect. The current implementation of TRRIP does not have temperature hints for data cache lines; data cache lines follow the default behavior of RRIP (lines 9-11 and 22-24).
Instruction cache lines without temperature information also follow default RRIP behavior.

Instruction accesses to the cache first requires address translation through the MMU where temperature information from the PTEs are obtained and then embedded in additional bits in the memory request. Note that the additional bits in the PTE and memory request already exist in many ARM-based mobile CPUs offering four implementation-defined bits~\cite{cortexa, cortexx}. As a result, we do not propose adding any additional bits and instead use at most two of the existing bits to encode the temperature information. TRRIP's replacement policy features only trigger on instruction memory requests containing valid temperature information. Otherwise, all cache lines are managed with the existing RRIP eviction mechanism regardless of their temperature. This is done to avoid storing the temperature of a cache line for its lifetime in the cache set, thus eliminating the need for additional bits in the cache set, or for the line's corresponding PTE to remain in the page table.

TRRIP can be implemented as two variants. Variant one is a minimal and only considers instruction cache lines classified as \textit{hot} -- most of the performance benefit from TRRIP should be obtained here as \textit{hot} code makes up the largest portion of executed instructions of an application. Variant two includes conditions for \textit{warm} and \textit{cold} instruction lines on top of variant one to assist keeping \textit{hot} lines in the cache for longer. 


\section{Evaluation}
\label{sec:evaluation}

\subsection{Simulator}
\label{sec:simulator}


\begin{table}[t]
\caption{Simulator configuration.}
        \footnotesize
\begin{tabular}{|c|l|}
\hline
\textbf{Component} & \textbf{Configuration}  \\ \hline \hline
\multirow{2}{*}{Core} & 6-wide dispatch, pseudo-FDIP prefetching,\\ 
                      & 128-entry ROB, 2GHz       \\ \hline
\multirow{3}{*}{Branch}  & 1K-entry BTB, 512-entry indirect-BTB,   \\
                      & 256-entry loop predictor, 1K-entry global predictor,   \\
                      & 8-cycle mispredict penalty \\ \hline    
\multirow{2}{*}{L1-D}  & 64KB, 4-way, LRU replacement, stride prefetcher,   \\
                      & 1/3 (tag/data)-cycle latency  \\ \hline    
\multirow{2}{*}{L1-I}  & 64KB, 4-way, LRU replacement, stride prefetcher,  \\
                      & 1/3 (tag/data)-cycle latency  \\ \hline 
\multirow{2}{*}{Unified Shared L2}  & 512kB (128kB per core), 8-way, TRRIP replacement,   \\
                      & inclusive, stride prefetcher, 8/12 (tag/data)-cycle latency  \\ \hline 
\multirow{2}{*}{Unified Shared SLC}  & 1MB, 16-way, LRU replacement, exclusive,  \\
                      & 10/30 (tag/data)-cycle latency       \\ \hline 
\multirow{2}{*}{DRAM}  & 8 chips/DIMM, 4 DIMMs, 7.6 GB/s controller bandwidth,      \\
                      & 400-cycle latency  \\ \hline 
\end{tabular}
\label{tab:simulator}
\end{table}

TRRIP is evaluated using Sniper~\cite{sniper}, a Pin-based~\cite{pin} simulator. Mobile OSs use energy aware scheduling and favor scheduling tasks to the energy-efficient cluster to save power even though they have less performant cores and memory. We simulate a core and memory hierarchy representative of an energy-efficient cluster on modern heterogeneous mobile SoCs (System-on-Chip) as they have the highest utilization~\cite{victor}. We model an out-of-order CPU with private 64kB L1-I and L1-D, a shared unified L2 cache, 128kB per core, or 512kB for the cluster assuming 4 cores in the cluster, and a shared unified SLC (System Level Cache)~\cite{victor}. Each cache contains stride-based hardware prefetchers. The simulation parameters are summarized in Table~\ref{tab:simulator}. We also implement FDIP-based~\cite{fdip} prefetcher which queries branch predictors to prefetch instructions ahead of instruction fetch. We do not model wrong-path execution/prefetching during branch mispredictions since the simulator is trace-based. Wrong-path prefetching can pollute the cache causing higher cache MPKI, thus the performance results in the following sections could be conservative. Our pseudo-FDIP based prefetcher gives an additional geomean performance improvement of 1.4\% on top of the stride-based (including next-line) prefetchers.

\subsection{Benchmarks \& Compilation Details}
\label{sec:benchmarks}


\begin{table}[t]
\caption{Benchmarks with data sets used for PGO profiles and performance measurement, and the amount of instructions fast-forwarded in the simulation.}
\centering
\footnotesize
\begin{tabular}{|l|c|c|c|}
\hline
\textbf{Benchmark}  & \textbf{Training} & \textbf{Evaluation} &\textbf{Fast Fwd.}\\ \hline \hline
abseil & all$^a$ & absl\_btree\_test & 1E9 \\ \hline
bullet & train$^b$ & eval$^b$ & 1E9 \\ \hline
clamscan & train$^b$ & eval$^b$ & 1E7 \\ \hline
clang  & ninja clang-check-c & gcc's ref & 1E8 \\ \hline
deepsjeng  & train$^b$ & ref$^b$ & 4E9 \\ \hline
gcc  & train$^b$ & ref$^b$ & 1E8 \\ \hline
omnetpp  & train$^b$ & ref$^b$ & 4E8 \\ \hline
python  & train$^b$ & test\_statistics & 1E8 \\ \hline
rapidjson   & unittest + perftest & perftest & 1E8 \\ \hline
\multirow{3}{*}{sqlite}  & --shrink-memory  & --shrink-memory &  \multirow{3}{*}{1E8}    \\
                      & --reprepare --size 50 & --reprepare --size 5 & \\
                      & --heap 10000000 64 & --heap 10000000 64 &   \\ \hline 
\end{tabular}

\footnotesize{$^a$ All tests in test suite.\\ $^b$ Use default profiling and evaluation input sets with benchmark.}
\label{tab:benchmarks}
\end{table}

A summary of the benchmarks and inputs for profiling and evaluation is listed in Table~\ref{tab:benchmarks}. We choose C/C++ based programs as mobile system software is primarily written in C/C++. Benchmarks \textit{clang}~\cite{clang}, \textit{gcc}~\cite{speccpu2017}, \textit{python}~\cite{cpython} represent the interpreters and JIT/AOT class of system components. Google's \textit{abseil}~\cite{abseil} represent calls to C++ libraries, \textit{bullet}~\cite{testsuite} for rendering, \textit{clamscan}~\cite{testsuite} for malware detection, Tencent's \textit{rapidjson}~\cite{rapidjson} for JSON parsing, and \textit{sqlite}~\cite{sqlite} for mobile database engine. We also choose additional integer type C++ benchmarks \textit{deepsjeng} and \textit{omnetpp} from CPU2017~\cite{speccpu2017} as mobile system software mostly contains integer data types. 
Different input sets are used for profile generation and evaluation.
All the benchmarks are built from source code using \textit{-O3} with full LTO (Link-Time Optimization) enabled and use LLVM's IR instrumentation PGO. Although ThinLTO~\cite{thinlto} is more widely adopted in industry due to lower compile time~\cite{bytedance-codesize, meta-merge, meta-reorder}, we chose full LTO to prove TRRIP is able to give performance improvement over highly optimized code. 
There are existing optimization passes that distribute code of the same function across code sections of different temperature (e.g., Hot-Cold Splitting~\cite{hotcoldsplit}, Machine Function Splitting~\cite{machinesplit}), but are disabled by default\footnote{Compiler optimization passes may be disabled by default due to functional or performance issues for specific application scenarios.}. TRRIP does not enable these features and uses the default PGO optimization pipeline.
Different benchmarks have different duration of start-up code which we are not interested in. We fast-forward to the hot part of the benchmark before starting performance measurements. The number of instructions that are fast-forwarded is listed for each benchmark in Table~\ref{tab:benchmarks}. All benchmarks are then simulated for 400M instructions after fast-forwarding.

\subsection{Evaluated Mechanisms}

\textbf{RRIP} (Re-Reference Interval Prediction)~\cite{rrip} has three variants, SRRIP (Static RRIP), BRRIP (Bimodal RRIP) and DRRIP (Dynamic RRIP). SRRIP is designed to resist cache misses due to \textit{scan} access patterns and BRRIP resists misses from \textit{thrashing} access patterns. As workloads are generally comprised of a mixture of different access patterns, DRRIP combines both SRRIP and BRRIP using set-dueling~\cite{set-dueling}.

\textbf{CLIP} (Code Line Preservation)~\cite{clip} is a RRIP-based policy that has preferential treatment of instruction cache lines for frontend bound applications. All instruction cache lines are inserted as \texttt{Immediate} re-reference. Set-dueling in CLIP may choose another variant of CLIP which prevents data cache lines from being promoted to \texttt{Immediate} re-reference like in baseline RRIP policies.

\textbf{SHiP} (Signature-based Hit Predictor)~\cite{ship} implements an additional predictor table on top of RRIP to improve re-reference prediction by correlating a signature to each cache line. Signatures can either be hashes of the memory address or PC. Signatures are used to predict if a line should be inserted as \texttt{Distant} re-reference to prevent polluting the cache. We implement a 64kB SHiP predictor at the L2 level and only apply it for instruction cache blocks using PC-based (same as memory address signatures for instruction memory) signatures.

\textbf{Emissary}\cite{emissary} identifies instruction cache lines that cause CPU frontend starvation in the decode and issue stages. Runtime metadata about cache lines likely to cause frontend starvation is passed to the L2 via additional bits per cache line. These instructions cache lines are prioritized in the unified L2 cache using way-locking built on top of LRU. The original Emissary work was completed on a different simulation infrastructure and evaluated using different workloads. We implement Emissary to the best of our ability on our infrastructure. We set 4 priority ways per set (in an 8-way set associative cache).

\textbf{TRRIP} cache replacement algorithm has two variants, TRRIP-1 and TRRIP-2. TRRIP-1 only considers \textit{hot}. TRRIP-2 considers \textit{hot}, \textit{warm} and \textit{cold} lines (see Section~\ref{sec:hardware}). 

We apply all policies to the L2 cache. All RRIP-based policies (i.e., \{S/B/D\}RRIP, CLIP, TRRIP) are modeled using 2-bit RRPV. Policies using set-dueling (i.e., DRRIP, CLIP) use 32 sampling sets for each policy, and use a 10-bit saturating counter (\texttt{PSEL}~\cite{set-dueling}) for policy decisions.

\subsection{Performance}
\label{sec:performance}
\begin{figure*}[!ht]
    \centering
    \includegraphics[width = 2\columnwidth]{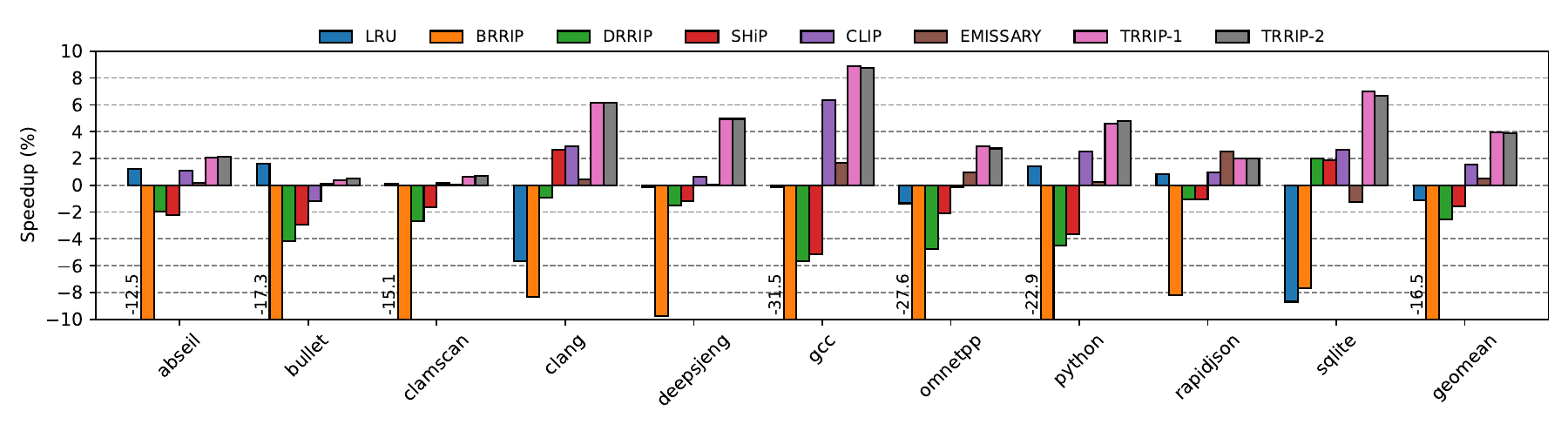}
    \caption{Speedup results of evaluated mechanisms are normalized to SRRIP replacement in the L2 cache.}
    \label{fig:performance-speedup}
\end{figure*}

\begin{table*}[!ht]
    \centering
    \caption{Raw L2 MPKI (Misses-Per-Kilo-Instruction) of SRRIP and L2 MPKI reduction for the evaluated mechanisms. Negative values means increased MPKI.}
    \footnotesize
    \begin{tabular}{|l|l|c|c|c|c|c|c|c|c|c|c|c|} \hline
        \multicolumn{2}{|c|}{\textbf{Benchmark}} & \textbf{abseil} & \textbf{bullet} & \textbf{clamscan} & \textbf{clang} & \textbf{deepsjeng} & \textbf{gcc} & \textbf{omentpp} & \textbf{python} & \textbf{rapidjson} & \textbf{sqlite} & \textbf{geomean} \\ \hline \hline
        \multirow{3}{*}{L2 MPKI} & Inst. & 1.79  & 0.13  & 0.36  & 16.68  & 0.7  & 3.54  & 4.71  & 4.83  & 0.57  & 4.08 & 1.67 \\
                              & Data & 17.52  & 1.76  & 2.73  & 19.51  & 1.22  & 5.99  & 12.3  & 11.04  & 8.36  & 6.99 & 6.28\\
                              & $\frac{\text{Inst.}}{\text{Data}}$ & 0.10 & 0.07 & 0.13 & 0.85 & 0.57 & 0.59 & 0.38 & 0.44 & 0.07 & 0.58 & 0.27\\ \hline \hline
        \multicolumn{13}{|c|}{\textbf{L2 MPKI Reduction (\%)}} \\ \hline \hline
        \multirow{2}{*}{LRU} & Inst. & 18.13  & 17.31  & 8.22  & -11.68  & -1.88  & -4.93  & -0.35  & 4.22  & 10.33  & -30.71 & 1.78\\
                                       & Data & 1.51  & -0.29  & -4.73  & 0.63  & -1.81  & 3.94  & -6.38  & 1.73  & 0.13  & -4.96 & -0.97 \\ \hline
        \multirow{2}{*}{BRRIP~\cite{rrip}} & Inst. & -225.93  & -205.47  & -115.31  & -20.86  & -52.57  & -105.7  & -88.54  & -106.54  & -135.36  & -4.17 & -94.54 \\
                                       & Data & -32.33  & -58.46  & -36.25  & -53.18  & -27.5  & -95.51  & -48.41  & -64.77  & -5.12  & -32.93 & -43.60 \\ \hline
        \multirow{2}{*}{DRRIP~\cite{set-dueling}} & Inst. & -25.81  & -26.85  & -15.74  & -2.3  & -8.34  & -11.39  & -11.44  & -12.9  & -16.74  & 11.18 & -11.52\\
                                       & Data & -3.08  & -7.36  & -3.85  & -4.96  & -3.11  & -8.78  & -5.28  & -6.49  & -0.66  & -5.88 & -4.92 \\ \hline
        \multirow{2}{*}{SHiP~\cite{ship}} & Inst. & -33.47  & -30.58  & -15.42  & 5.85  & -6.08  & -8.43  & -7.27  & -12.61  & -14.22  & 7.15 & -10.81\\
                                       & Data & -3.94  & -6.7  & -1.2  & -5.05  & -0.84  & -12.18  & 0.56  & -7.31  & 1.26  & -0.89 & -3.55 \\ \hline
        \multirow{2}{*}{CLIP~\cite{clip}} & Inst. & 18.92  & 26.45  & 14.32  & 5.4  & 9.7  & 16.91  & 9.68  & 10.89  & 12.96  & 8.76 & 13.6 \\
                                       & Data & 1.03  & -9.27  & -5.66  & -2.87  & -7.04  & -0.75  & -9.88  & -0.38  & -0.1  & -7.33 & -4.15 \\ \hline
        \multirow{2}{*}{Emissary~\cite{emissary}} & Inst. & 7.83  & 38.65  & 10.09  & 5.94  & -4.16  & 27.16  & 14.19  & 0.95  & 68.71  & -27.16 & 22.06 \\
                                            & Data & 0.43  & -17.74  & -0.49  & -15.94  & 1.69  & -8.45  & -2.48  & 0.63  & 1.38  & -4.43 & -4.33\\ \hline
        \multirow{2}{*}{TRRIP-1} & Inst. & 37.88  & 22.92  & 10.35  & 5.92  & 47.36  & 27.2  & 14.3  & 22.97  & 43.22  & 20.58 & 26.49 \\
                                            & Data & 0.42  & -0.49  & -0.65  & -15.73  & -9.24  & -8.29  & -2.38  & -5.57  & 0.54  & -8.73 & -4.89 \\ \hline
        \multirow{2}{*}{TRRIP-2} & Inst. & 39.39  & 26.79  & 11.08  & 5.96  & 47.36  & 26.93  & 17.3  & 23.29  & 43.22  & 19.38 & 27.26\\
                                            & Data & 0.44  & -0.72  & -0.71  & -15.82  & -9.25  & -8.92  & -5.09  & -6.2  & 0.54  & -9.35 & -5.38 \\ \hline
    \end{tabular}
    \label{tab:L2_MPKI}
\end{table*}

Figure~\ref{fig:performance-speedup} plots the speedup, measured as the reduction in CPU execution cycles for a fixed number of instructions (i.e., 400M). Table~\ref{tab:L2_MPKI} lists raw MPKI and MPKI reductions for instruction and data lines for the evaluated benchmarks. All measurements for the various techniques are normalized to SRRIP running benchmarks optimized using PGO. RRIP-based policies prove to be a good baseline to further optimize as the baseline SRRIP has higher speedup compared to LRU. All of the evaluated benchmarks show higher data MPKI compared to instruction MPKI in the L2 cache, however almost all techniques that prioritize instruction cache lines over data lines, namely CLIP, Emissary, and TRRIP, improve performance. Instruction MPKI is reduced while data MPKI increases slightly. The tradeoff is profitable as modern cores are able to hide backend memory stalls through superscalar and out-of-order execution, where instruction misses cause frontend stalls which are difficult to conceal.   
Both variants of TRRIP outperform the evaluated prior art, especially on benchmarks with higher ratio of instruction MPKI to data MPKI, i.e., \textit{clang}, \textit{gcc}, \textit{sqlite}, \textit{deepsjeng} and \textit{python}. \textit{Deepsjeng}'s raw instruction MPKI number is relatively low compared to the other benchmarks; the significant MPKI reduction (over 47\%) leads to the 5.0\% performance improvement in both variants of TRRIP.
TRRIP-2 inserts warm instruction lines at \texttt{Near} re-reference to avoid displacing \textit{hot} cache lines while still prioritizing \textit{warm} cache lines over data lines. This results in higher instruction MPKI reduction in TRRIP-2 (geomean 27.3\%) compared to TRRIP-1 (geomean 26.5\%). Overall the minute reduction of instruction MPKI in TRRIP-2 does not translate to noticeable performance gains as warm code is executed considerably less than hot code, and has less coverage of costly instruction cache misses explored in Section~\ref{sec:eval-costly-coverage}. Across all benchmarks, TRRIP obtains geomean performance improvement of 3.9\% over SRRIP for both variants by reducing instruction MPKI by 26.5\% and 27.3\% for TRRIP-1 and TRRIP-2 respectively.

Emissary prioritizes instruction cache lines that cause frontend stalls through decode starvation and is implemented on LRU, and provides speedup of 0.5\% over SRRIP, and 1.6\% relative to baseline LRU. Benchmarks \textit{bullet} and \textit{rapidjson} show the highest instruction MPKI reduction using Emissary because there are areas of code not covered by TRRIP -- further explored in Section~\ref{sec:eval-costly-coverage}.
CLIP prioritizes all instruction caches in its replacement policy and does not rely on a compiler marking hot code. In benchmarks \textit{bullet} and \textit{clamscan} where cache misses may land in code unreachable by TRRIP, CLIP can provide higher MPKI reduction similar to Emissary. CLIP provides a performance improvement of 1.6\% over SRRIP. Section~\ref{sec:eval-hot-threshold} details how TRRIP's preference of hot code is more advantageous than treating all instruction cache lines as high priority in CLIP. 
Our implementation of SHiP predicts if instructions lines have \texttt{Distant} re-reference and inserts such lines at the lowest priority to prevent pollution and shows speedup over 1.9\% compared to SRRIP in \textit{sqlite}, but on average does not perform well since the rest of the applications do not exhibit memory accesses patterns that SHiP aims to optimize. Set-duelling based DRRIP also does not perform well since 32 of the sampling sets use BRRIP. The evaluated benchmarks may not show thrashing access pattern BRRIP is intended to optimize for, thus DRRIP's performance suffers as expected~\cite{rrip}.

\subsection{Power and Area}
\label{sec:power-area}
\begin{table}[!t]
\caption{Static power and area overheads.}
\centering
\footnotesize
\begin{tabular}{|l|r|r|}
\hline
\textbf{Mechanism} & \textbf{Static Power (\%)}  & \textbf{Area (\%)} \\ \hline \hline
TRRIP & $\sim$0.0 & $\sim$0.0 \\ \hline
CLIP~\cite{clip} & $\sim$0.0 & $\sim$0.0 \\ \hline
Emissary~\cite{emissary} & 0.5 & 0.7 \\ \hline
SHiP~\cite{ship} & 1.7 & 3.0 \\ \hline

\end{tabular}
\label{tab:power-area}
\end{table}

We use McPAT~\cite{mcpat} with 22nm technology node to simulate static power and area for the state-of-the-art evaluated techniques, TRRIP, SHiP, CLIP, and Emissary. We only model the power and area of the on-chip components (i.e., core, L1-D, L1-I, and L2); the last level (e.g., SLC) cache is usually off-chip in modern CPUs. Both SHiP and Emissary require microarchitecture changes to CPU frontend. PC signatures need to be propagated through the I-TLB in SHiP, and Emissary requires additional logic to track and report frontend starvation to memory. These changes are too complicated to model accurately in McPAT so we optimistically assume they do not increase core power and area. TRRIP and CLIP requires no modifications to the CPU microarchitecture.
Table~\ref{tab:power-area} summarize the static power and area overheads of the techniques, normalized to SRRIP implementation. SHiP has the highest power and area overheads due to the additional 64kB predictor. Emissary has lower overheads since no additional storage structures are needed, other than additional two bits in the L1 and L2 cache lines.
Additional bits to store TRRIP's temperature information in the PTEs and mechanism to transfer the temperature information with memory requests is already implemented in commercial processors used in mobile systems (e.g., ARM PBHA)~\cite{cortexa, cortexx}, therefore is not included as an overhead since no new storage is needed.
TRRIP is able to give performance on-top of PGO almost for free and no additional storage is needed compared the prior two techniques.

\subsection{Coverage of Costly Instruction Misses}
\label{sec:eval-costly-coverage}

\begin{figure}[!t]
    \centering
    \begin{subfigure}[t]{1\columnwidth}
        \centering
        \includegraphics[width = 1\columnwidth]{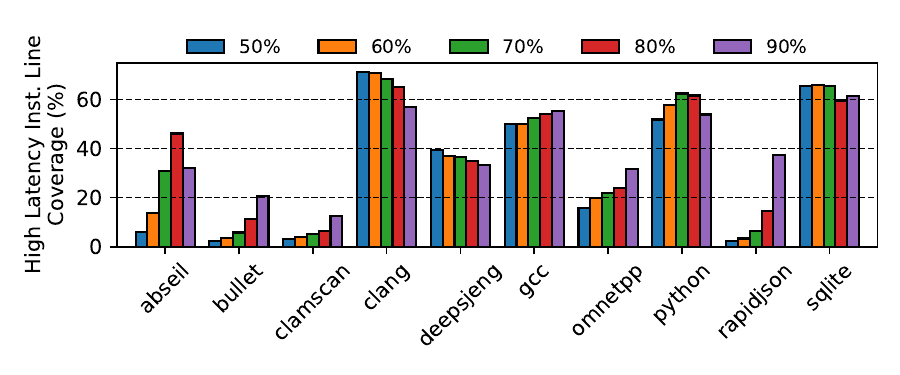}
        \caption{Coverage within TRRIP's \textit{hot} text section.}
        \label{fig:high-latency-coverage}
    \end{subfigure}
    
    \begin{subfigure}[t]{1\columnwidth}
        \centering
        \includegraphics[width = 1\columnwidth]{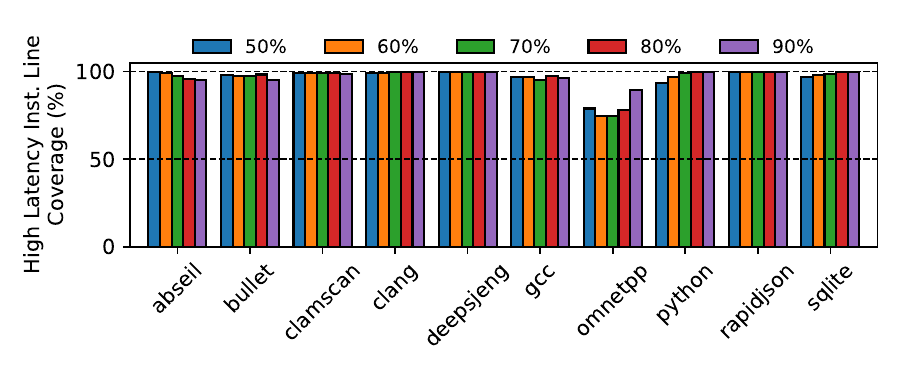}
        \caption{Coverage within TRRIP's \textit{hot} text section, excluding misses in external code.}
        \label{fig:high-latency-coverage-filtered}
    \end{subfigure}
\caption{Coverage of costly instruction cache misses in TRRIP for different top $\text{N}^{\text{th}}$ percentile of costly instruction misses.}
\end{figure}

Emissary~\cite{emissary} postures that some instruction misses are more costly than others. Costly instruction cache lines are those that incur a cache miss and cause decode starvation, and also contain instructions that retire (not mispredicted). Emissary prioritizes these costly caches lines in the instruction caches. 
Emissary's mechanism requires additional bits per cache line to store this information. TRRIP conversely uses execution frequency of code (i.e., temperature) to prioritize instruction lines in the cache which does not require additional storage, but it is unable to distinguish which instruction misses are more costly.
Figure~\ref{fig:high-latency-coverage} shows the percent coverage of the top $\text{N}^{\text{th}}$ percentile of costly instruction misses seen during execution of hot code identified by TRRIP. The hot code sections in \textit{bullet}, \textit{clamscan}, \textit{omentpp}, \textit{rapidjson} show low coverage of costly instruction misses which explains the lower performance improvement. This is because many of the costly instruction misses fall either in the PLTs (Procedure Linkage Tables) or external code (i.e., other libraries) that are not transparent in TRRIP's compile phase. In theory, costly instruction misses landing in the PLT can be handled through linker-based optimizations, such as incorporating the PLT section with TRRIP's hot code section; this exploration is left for future work. 
Pure hardware techniques (e.g., Emissary and CLIP) are advantageous as they work on all executed code, and are not limited to what the compiler compiles. However, the trade-off is that pure hardware techniques require extra storage or are less flexible to disable if performance degradation is observed.

Figure~\ref{fig:high-latency-coverage-filtered} shows the top $\text{N}^{\text{th}}$ percentile of costly instruction misses limited to TRRIP's compiled code (which excludes PLTs and external libraries). Nearly all of the costly instruction misses are on hot code, showing TRRIP is still able to prioritize costly instruction misses without the hardware overheads of Emissary by relying on practical offline software analysis. ~\textit{Omnetpp}'s lower coverage is due to some of the costly instruction misses landing in warm code.

TRRIP's coverage can be improved by compiling external dependencies with TRRIP's PGO. Many of the most commonly used mobile system software components/libraries are optimized using PGO, thus TRRIP's code coverage would be higher if deployed on mobile platforms.

\subsection{Sensitivity to Hot Code Selection}
\label{sec:eval-hot-threshold}

\begin{equation}
C_{threshold} = C_{total} \times Percentile_{hot}
\label{eq:pgo-count-threshold}
\end{equation}


\begin{equation}
\exists C_n : C_n = C_i : C_i > C_{i+1}\: \text{and}\: \sum_{i=0}^{n}C_i < C_{threshold}
\label{eq:pgo-hot-threshold}
\end{equation}

\begin{figure}[!t]
    \centering
    \begin{subfigure}[!t]{1\columnwidth}
        \centering
        \includegraphics[width = 1\columnwidth]{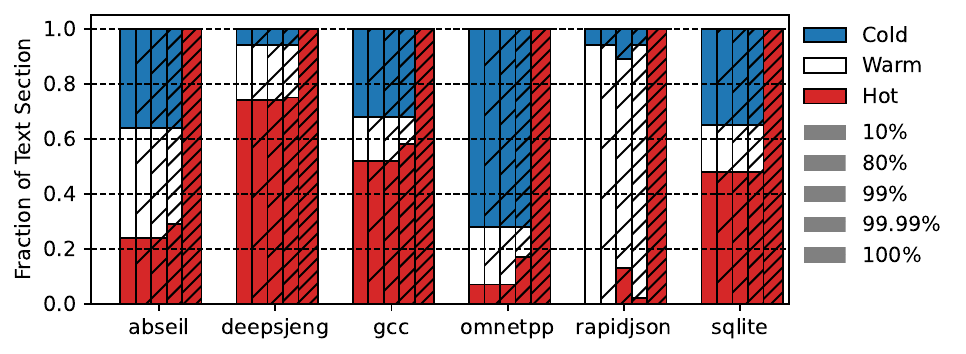}
        \caption{Distribution of \textit{hot}, \textit{warm} and \textit{cold} text sections.}
        \label{fig:hot-code-fraction}
    \end{subfigure}
    
    \begin{subfigure}[!t]{1\columnwidth}
        \centering
        \includegraphics[width = 1\columnwidth]{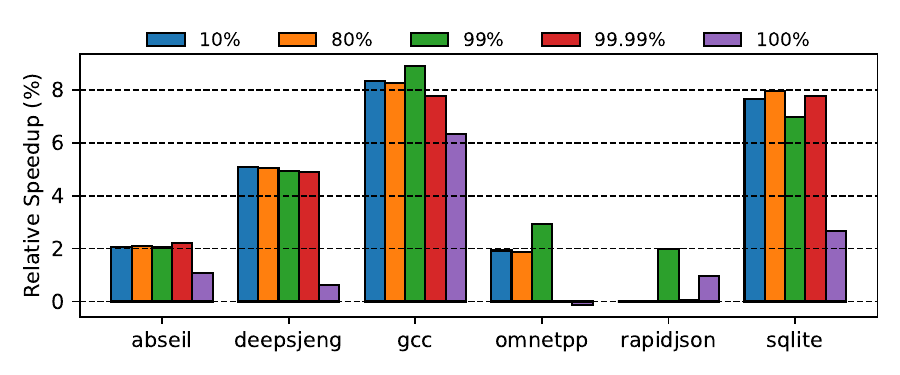}
        \caption{Performance gain.}
        \label{fig:hot-threshold-sensitivity-s1}
    \end{subfigure}
\caption{Sensitivity to different compiler hot thresholds ranging from 10\% to 100\%.}
\end{figure}

Compiler PGO determines code temperature based on thresholds calculated from BB counters in the profiling data. Equations~\ref{eq:pgo-count-threshold} and~\ref{eq:pgo-hot-threshold} describe the compiler's logic to classify \textit{hot} code. A similar logic is also used to classify \textit{cold} code. Code is considered as \textit{hot} if its BB counter is greater than a calculated counter value $C_{n}$.  $C_{n}$ is determined from Equation~\ref{eq:pgo-hot-threshold} where BB counters ($C_{i}$) are first sorted highest to lowest and summed until  a threshold ($C_{threshold}$) is exceeded; $C_{n}$ is the count prior to exceeding $C_{threshold}$. $C_{threshold}$ is a percentile ($Percentile_{hot}$) of the total counts ($C_{total}$) in the profile from Equation~\ref{eq:pgo-count-threshold}. $Percentile_{hot}$ is a compile-time adjustable value where the closer to 100\% it is, the more code will be considered as \textit{hot}.

Figure~\ref{fig:hot-code-fraction} shows the fraction of \textit{hot}, \textit{warm}, and \textit{cold} text sections relative to the entire text section\footnote{Smaller text sections (e.g., \texttt{text.init}, \texttt{text.fini} are not included).} using different $Percentile_{hot}$ for a select subset of the benchmarks. \textit{Hot} text section size does not noticeably increase until the hot threshold $Percentile_{hot}$ is increased beyond 99\% due to the nature of the distribution of $C_{n}$ values in the profile\footnote{The distribution $C_{n}$ shows a long tail of warm and cold blocks, thus it takes a relatively high $Percentile_{hot}$ value to see the distinction between hot and non-hot blocks change. }. The open-source LLVM compiler uses $Percentile_{hot}=99\%$ by default. Setting $Percentile_{hot}$ to 100\% is similar to CLIP. 

Figure~\ref{fig:hot-threshold-sensitivity-s1} shows the performance gain of TRRIP-S1 when setting $Percentile_{hot}$ from 10\% to 100\%. We rebuild the application each time the $Percentile_{hot}$ value is changed. The speedup numbers are normalized to the same executable run with baseline SRRIP policy. TRRIP prioritizes more instruction cache lines as $Percentile_{hot}$ increases, up to 100\% in which case all instruction cache lines are considered \textit{hot}. Different applications respond best to different $Percentile_{hot}$ values, however the key observation is that being selective about prioritizing cache lines is needed to maximize performance gain. Prioritizing all instruction lines ($Percentile_{hot}=100\%$), similar to CLIP still obtains performance gains in most cases, but is substantially less than being selective. TRRIP only prioritizes instruction cache lines selectively classified as \textit{hot} via PGO analysis to maximize performance benefit.

\subsection{Impact of Cache Size and Associativity}
\label{sec:cache-size}
\begin{figure}[!t]
    \centering
    \begin{subfigure}[t]{1\columnwidth}
        \centering
        \includegraphics[width = 1\columnwidth]{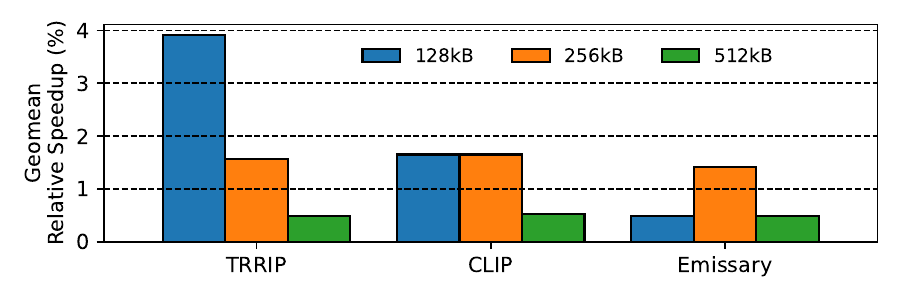}
        \caption{TRRIP-1, CLIP and Emissary on 128kB, 256kB, 512kB L2 with 8-way set associativity.}
        \label{fig:cache-size-sensitivity}
    \end{subfigure}
    
    \begin{subfigure}[t]{1\columnwidth}
        \centering
        \includegraphics[width = 1\columnwidth]{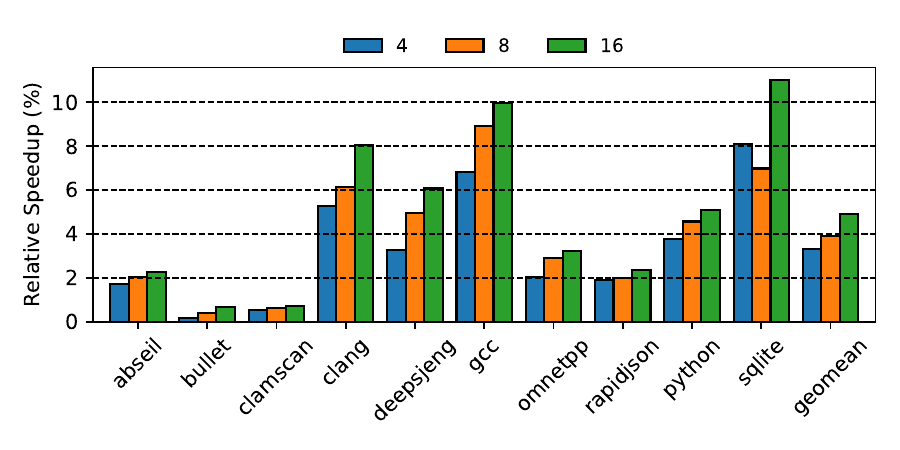}
        \caption{TRRIP-1 on 4-way, 8-way, 16-way set associativity on 128kB L2.}
        \label{fig:cache-assoc-sensitivity}
    \end{subfigure}
\caption{Cache size and associativity sensitivity.}
\end{figure}

Figure~\ref{fig:cache-size-sensitivity} shows the sensitivity to cache size for TRRIP-1, CLIP and Emissary. Only the geomean is displayed for all 10 benchmarks due to space. Larger cache sizes suffer less from cache misses and hence see less relative performance improvement from cache replacement optimizations. TRRIP's limitation is that hot code (and warm code in TRRIP-2) is only prioritized by TRRIP's replacement policy if the code is compiled using TRRIP's PGO compiler. Applications that spend many cycles executing code outside of the scope of the TRRIP compiled binary will have low coverage of prioritizing instruction lines that are both hot and expensive to miss on as seen in the previous Subsection~\ref{sec:eval-costly-coverage}. If TRRIP's dynamic code footprint coverage of hot/warm code fits within the cache then less performance gain is expected. Pure hardware techniques like CLIP and Emissary do not have this limitation, which explains the smaller reduction in performance gain relative to TRRIP as cache size increases. TRRIP's gain is expected to increase if the application's external code is also compiled using TRRIP's compiler PGO. Note that the power and area overheads of prior hardware based solutions (e.g., SHiP, Emissary) scales with cache size increases and will continue to consume additional power even in situations where the frontend is not a bottleneck. TRRIP adds negligible hardware overheads regardless of whether or not the workload experiences frontend bottlenecks, and is easily disabled compared to hardware-only replacement policies (i.e., CLIP).

Figure~\ref{fig:cache-assoc-sensitivity} shows cache associativity sensitivity of TRRIP on 4-way, 8-way and 16-way set associative L2. TRRIP consistently obtains more performance improvement with higher set associativity as hot cache lines showing larger reuse distances (seen in Figure~\ref{fig:l2-reuse-cdf}) are captured more effectively. 

Since TRRIP does not require additional space in the cache, it can be extended across the cache hierarchy, e.g., to the L1 instruction cache. Targeting more caches and improving TRRIP's code coverage can promote adoption in server-class systems which have larger cache sizes and code footprints compared to mobile systems. We leave this exploration for future work.

\subsection{Using Larger Pages}
\label{sec:large-pages}

\begin{table}[!t]
\caption{Pages used (hot/warm) and binary size.}
\centering
\footnotesize
\begin{tabular}{|l|r|r|r|r|}
\hline
\textbf{Benchmark} & \textbf{4kB pages} & \textbf{16kB pages}  & \textbf{2MB pages} & \textbf{Binary Size (B)} \\ \hline \hline
abseil & 142/244 & 36/61 & 1/1 & 6.1M \\ \hline
bullet & 29/26 & 8/7 & 1/1 & 915K\\ \hline
clamscan & 15/24 & 4/6 & 1/1 & 732K\\ \hline
clang & 2153/2376 & 539/594 & 5/5 & 168M\\ \hline
deepsjeng & 25/6 & 7/2 & 1/1 & 164K\\ \hline
gcc & 1334/463 & 334/116 & 3/1 & 14M\\ \hline
omnetpp & 31/99 & 8/25 & 1/1 & 3.3M\\ \hline
python & 175/282 & 44/71 & 1/1 & 22M \\ \hline
rapidjson & 79/464 & 20/116 & 1/1 & 8.4M\\ \hline
sqlite & 136/48 & 34/12 & 1/1 & 1.5M\\ \hline

\end{tabular}
\label{tab:page-util}
\end{table}

Page sizes on both mobile and server platforms are typically 4kB by default. 
Mobile platforms recently added support of 16kB pages (e.g., AOSP 15~\cite{aosp-page}). 
Server-class systems can typically scale up to 2MB pages. 
Table~\ref{tab:page-util} documents the number of pages used for \textit{hot} and \textit{warm} text sections for 4kB, 16kB and 2MB page sizes, rounded up to the nearest full page. 
The codes of different temperatures are placed in their respective code sections in the ELF (Figure~\ref{fig:elf-layout}); code of one temperature (i.e., \textit{cold}) is not placed in the code section of a different temperature (i.e., \textit{ hot}). Most pages will contain a single temperature code, especially for smaller pages sizes, i.e., 4kB or 16kB mobile pages. 

A page may contain code of different temperatures only if it overlaps two text sections of different temperatures; marking such overlapping pages with a single temperature potentially risks TRRIP to prioritize lines incorrectly (e.g., treating \textit{warm} as \textit{hot}).
This problem can increase as larger pages are used to improve TLB span.
Prevention mechanisms of inaccurate page temperatures can include: (1) adding padding between different temperature code sections so that they never reside on the same page at the cost of using additional page space, (2) disable marking pages that overlap code sections of different temperatures with temperature information, (3) keep code pages small to improve temperature accuracy and allow other pages to be large (mixed page sizes are supported in Hugepage~\cite{hugepages}). TRRIP can be disabled to restore performance if regression is observed when applying aforementioned prevention mechanisms.
\section{Related Work}
\label{sec:related-work}
\subsection{Replacement Policies}

RRIP~\cite{rrip} has gained popularity due to its performance advantage over traditional techniques (i.e., LRU) in addition to having lower hardware complexity. As a result, many state-of-the-art cache policies are built on top of RRIP. Hawkeye~\cite{hawkeye}, Mockingjay~\cite{mockingjay} and SHiP~\cite{ship} aim to predict future memory access patterns to mimic the behavior of Belady's optimal replacement~\cite{belady}; all require additional hardware to make these predictions. 
CLIP~\cite{clip}, also based on RRIP, has negligible hardware implementation overheads, however it blindly treats all instruction cache lines as the same.
Our evaluation shows TRRIP's temperature-based classification of instruction cache lines can yield additional performance improvement over CLIP. PDP~\cite{pdp} and Emissary~\cite{emissary} protects cache lines form being prematurely evicted by locking them into the cache for a certain duration. Both these techniques need additional hardware, either in the form of predictors as in PDP, or additional bits per cache line in Emissary which adds storage overheads. Thermometer~\cite{thermometer} uses PGO to identify \textit{hot} branches based on a \textit{hit-to-taken} ratio to modify BTB (Branch Target Buffer) replacement, requiring additional storage in the BTB and modifications to the ISA.

TRRIP in contrast to the aforementioned approaches does not require additional hardware or ISA changes. The selection of instruction cache lines to prioritize in the caches is done offline by the compiler with PGO. TRRIP features are also easy to toggle on and off due to the nature of it being a co-designed technique unlike pure hardware mechanism (e.g., CLIP).

\subsection{Instruction Prefetching}
Prefetching instruction cache lines is an alternate approach to reduce CPU front-end stalls. Conventional software instruction prefetching is notorious for being finicky -- timing and accuracy of the software prefetches is difficult to get correct both from programmer inserted prefetches, or prefetches inserted by the compiler. Prior work~\cite{ispy, asmdb, twig} has made inroads into solving issues with software instruction prefetching, such as using PGO to improve accuracy; however, these techniques require inserting new instructions which incur execution overheads that diminish the gains from reducing instruction cache misses.

Runahead instruction prefetchers~\cite{fdip, pdip} are hardware mechanisms that query the branch predictor units to prefetch the future path of the program. These techniques are more adaptable to different runtime behaviors (unlike statically inserted software prefetches), but the problem of ever-growing code footprint is overwhelming branch predictor capacity~\cite{llbp}, and hence adversely affects prefetcher accuracy.
Record-and-replay~\cite{pif,mana} techniques record instruction sequences and replay them as prefetches at a later time, and co-designed methods embed offline analysis metadata into the binary guide the prefetcher~\cite{hierarchical-prefetch} to workaround large code footprint issues; however, require additional logic and storage requirements on and off-chip.

TRRIP relies on PGO similar to existing software prefetching techniques, but does not insert any new instructions into the program. TRRIP also has negligible power and area overheads unlike aggressive hardware instruction prefetching techniques~\cite{pif,mana,hierarchical-prefetch}. Software/hardware prefetching are orthogonal methods to TRRIP, and can be used in conjunction to further reduce frontend bottlenecks.
\section{Conclusion \& Future Work}
\label{sec:conlcusion}

Modern mobile workloads have complex runtime behaviors and large instruction footprints which cause CPU frontend stalls due to instruction caching misses. Increasing cache sizes or implementing complicated hardware is not a panacea to the problem due to on-chip power and area constraints. An almost-free solution we propose to tackle this problem is TRRIP, a software-hardware codesign mechanism leveraging PGO to classify code \textit{temperature}, mainly identification of the most executed (\textit{hot}) instructions to prioritize keeping in the cache. Lightweight software to hardware interfaces passes code temperature information down to the cache hierarchy to trigger code temperature based behaviors in the cache replacement policy requiring very little modification to the hardware. We show that TRRIP can reduce L2 instruction MPKI by 26.5\% across our sampled benchmarks yielding geomean speedup of 3.9\% over SRRIP cache replacement running applications with PGO applied.

In this paper we apply TRRIP to instruction cache replacement policies to reduce CPU frontend stalls by preventing cache misses on the \textit{hottest}, most executed code. However, the ever-growing code complexity and footprint of modern workloads does not only overwhelm caches, it is a problem for all on-chip features which require on-chip storage and replacement policies. In our future work, we plan to apply TRRIP to other hardware where large and complicated code will be a problem, such as the branch prediction units (e.g., BTB) and TLB. TRRIP can be extended to cover data cache lines which will require new compiler analysis and transformations.
TRRIP may also benefit server-class applications and processors, which may require implementation into different compiler infrastructures (e.g., JIT compilation). Coarse-granularity code temperature information may also be learned dynamically to improve code coverage for TRRIP.


\bibliographystyle{ACM-Reference-Format}
\bibliography{refs}

\end{document}